\documentclass[aps,prl,twocolumn,superscriptaddress]{revtex4}

\usepackage{epsf,epsfig,amsmath,amssymb}

\newsavebox{\ns}
\newsavebox{\dbrane}

\def\be{\begin{equation}}
\def\ee{\end{equation}}
\def\bea{\begin{eqnarray}}
\def\eea{\end{eqnarray}}

\def\Dslash{\,\,{\raise.15ex\hbox{/}\mkern-12mu D}}
\def\Dbarslash{\,\,{\raise.15ex\hbox{/}\mkern-12mu {\bar D}}}
\def\delslash{\,\,{\raise.15ex\hbox{/}\mkern-9mu \partial}}
\def\delbarslash{\,\,{\raise.15ex\hbox{/}\mkern-9mu {\bar\partial}}}
\def\pslash{\,\,{\raise.15ex\hbox{/}\mkern-9mu p}}
\def\calDslash{\,\,{\raise.15ex\hbox{/}\mkern-12mu {\cal D}}}

\newcommand\diff{\mbox{d}}

\newcommand{\de}{\partial}
\newcommand{\vol}{\mbox{vol}}

\newcommand{\reef}[1]{(\ref{#1})}

\newcommand{\dd}{\diff}

\newcommand{\bbR}{\mathbb{R}}

\newcommand{\me}{\mbox{e}}

\begin{document}

\title{The central charge of supersymmetric $AdS$$_5$ solutions of type
  IIB supergravity} 

\author{Maxime Gabella}
\affiliation{Rudolf Peierls Centre for Theoretical Physics, University of
  Oxford, 1 Keble Road, Oxford OX1 3NP, U.K.} 
\author{Jerome P. Gauntlett}
\affiliation{Theoretical Physics Group, Blackett Laboratory,
  Imperial College, London SW7 2AZ, U.K.}
\affiliation{The Institute for Mathematical Sciences,
  Imperial College, London SW7 2PE, U.K.}
\author{Eran Palti}
\affiliation{Rudolf Peierls Centre for Theoretical Physics, University of
  Oxford, 1 Keble Road, Oxford OX1 3NP, U.K.} 
\author{James Sparks}
\affiliation{Mathematical Institute, University of Oxford,
  24-29 St Giles', Oxford OX1 3LB, U.K.}
\author{Daniel Waldram}
\affiliation{Theoretical Physics Group, Blackett Laboratory,
  Imperial College, London SW7 2AZ, U.K.}
\affiliation{The Institute for Mathematical Sciences,
  Imperial College, London SW7 2PE, U.K.}

\begin{abstract}
We show that generic supersymmetric $AdS_5$ solutions of type IIB
supergravity admit a canonical contact structure.  This structure
determines the central charge
of the dual field theory and the conformal dimension of operators
dual to supersymmetric wrapped D3-branes. 
Hence both quantities can be calculated using incomplete information about the
solutions, allowing us to prove that they are rational
numbers for solutions with a $U(1)$ $R$-symmetry, in
agreement with field theory expectations. We also discuss related
Duistermaat-Heckman integrals and localization formulae. 
\end{abstract} 

\maketitle

\section{Introduction}

In its simplest setting, the AdS/CFT correspondence is an equivalence
between $d$-dimensional conformally invariant quantum field theory
(CFT) and string theory (or M-theory) in a ten- (or eleven-)
dimensional background of the form $AdS_{d+1}\times Y$, where
$AdS_{d+1}$ is $(d+1)$-dimensional anti-de Sitter space and $Y$ is a
compact, internal manifold. Generically, the weakly coupled limit of
the string/M-theory background describes a strongly coupled limit of 
the CFT. This remarkable duality has provided powerful new insights
into strong coupling phenomena, with potential 
applications to condensed matter and nuclear physics systems. 

One focus has been to study superconformal field theories (SCFTs) in
four dimensions with ${\cal N}=1$ supersymmetry. Such SCFTs
are invariant under an abelian $R$-symmetry which encodes
considerable information about the theory. For example, the anomalous
dimensions of a special class of operators ${\cal O}$, known as chiral
and anti-chiral primaries, are given, exactly, by their $R$-charges
via $\Delta({\cal O})=\tfrac{3}{2}|R({\cal O})|$.
Furthermore, the central charges, which determine the conformal
anomaly, are fixed by the $R$-symmetry via~\cite{freedman}
\bea\label{aandc}
a = \tfrac{3}{32}\left(3\mbox{Tr} R^3 - \mbox{Tr}R\right),\quad
c = \tfrac{1}{32}\left(9\mbox{Tr} R^3 - 5\mbox{Tr}R\right)
\eea
where the trace is over the $R$-charges of the fermions in the theory.

An interesting subclass of such SCFTs have a dual description in terms
of type IIB string backgrounds of the form $AdS_5\times Y$. 
In the supergravity limit, requiring the background to be
supersymmetric implies particular restrictions on the geometry of
$Y$~\cite{Gauntlett:2005ww}. The dimension of certain chiral operators
of the SCFT can then be deduced via Kaluza-Klein (KK) reduction of the
supergravity modes on $Y$, while others can be computed by analysing
supersymmetric sub-manifolds of $Y$ on which branes can wrap. In this
limit, the SCFTs necessarily have $a=c$, the value of which may be
calculated by determining Newton's constant, $G_5$, in the effective
five-dimensional KK reduced theory~\cite{kostas}. 

The $R$-symmetry of the SCFT corresponds to a Killing vector field
$\xi$ on $Y$, which has a canonical construction given the background
is supersymmetric~\cite{Gauntlett:2005ww}.  If the orbits of this
Killing vector field all close the $R$-symmetry is $U(1)$, while if
they do not it is $\bbR$. In light of the comments above, one would
like to understand in general how the $R$-symmetry Killing vector can
be used to calculate quantities of interest for the dual SCFT, in
particular, $a$ and $\Delta({\cal O})$.  Indeed, the $a$ central
charge is an important quantity in a $d=4$ CFT, being (conjecturally)
a count of the massless degrees of freedom in the theory.

To date, the most widely studied class of dual solutions are in a 
particular class where only the metric and the self-dual five-form
flux $F_5$ in the supergravity theory are non-vanishing. In this case $Y$ is
necessarily a Sasaki-Einstein manifold $SE_5$. By definition, the
metric cone over $SE_5$ is Calabi-Yau and the $AdS_5$ solutions
arise from the near-horizon limit of D3-branes sitting at the apex of 
this cone. The central charge is inversely proportional to the volume of
$SE_5$~\cite{gubser}. In this case, $\xi$ is known as the Reeb vector
field, and using the Sasaki-Einstein geometry, several formulae for
$a$ and $\Delta({\cal O})$ in terms of $\xi$ can indeed be
derived~\cite{MSY1,MSY2}. These formulae are significant because they
allow one to calculate important physical quantities without requiring
the full explicit solution. Although rich classes of explicit $SE_5$
solutions are known, their construction relies on the solutions having
a large amount of symmetry and they comprise a small subset of all
solutions. 

In this letter we show~\footnote{A more detailed discussion of
  some of the calculations here will appear in~\cite{GCGpaper}.} 
that such formulae still arise in the case of generic supersymmetric
$AdS_5$ type IIB solutions. This is rather remarkable since the
derivations in \cite{MSY1,MSY2} for the $SE_5$ case used the
Calabi-Yau property of the cone which no longer holds in general. Our
analysis only excludes the special case when $F_5$
vanishes. Physically, this means we require that the D3-brane charge
is non-vanishing and this includes all known $AdS_5$ type IIB
solutions.

Our key observation is that, remarkably, supersymmetry implies that
these solutions all have a canonical contact structure. 
A contact structure on $Y$ is a one-form
$\sigma$ such that the five-form $\sigma\wedge \dd\sigma\wedge
\dd\sigma$ is nowhere-vanishing. Equivalently there is a natural
symplectic structure on the cone over $Y$. We will find that, up to a
universal constant of proportionality, the central charge $a$ is
the ``contact volume'' of $Y$. In addition, the conformal
dimension $\Delta({\mathcal O}_{\Sigma_3})$ of the chiral primary
operator ${\mathcal O}_{\Sigma_3}$ dual to a D3-brane wrapped on a
supersymmetric submanifold $\Sigma_3\subset Y$ is given by the contact
volume of $\Sigma_3$. More precisely,
\bea\label{centralchargesigma}
\frac{a_{{\cal N}=4}}{a} &=& \frac{1}{(2\pi)^3}
\int_Y{ \sigma \wedge \dd\sigma \wedge \dd\sigma}~,\\
\label{conformal}
\Delta({\mathcal O}_{\Sigma_3})& = &\frac{2\pi N\int_{\Sigma_3}\sigma\wedge\dd \sigma}{\int_Y\sigma\wedge
\dd\sigma\wedge\dd\sigma}~,
\eea
where $a_{{\cal N}=4}=N^2/4$ is the (large $N$) central charge for
$SU(N)$ ${\cal N}=4$ super Yang-Mills theory, and $N$ is the quantized
D3-brane charge. We expect that the existence of this contact
structure will be important in extending other results from the $SE_5$
case to generic $AdS_5$ solutions.  

\section{Supersymmetric type IIB $AdS_5$ solutions}

Our main results follow directly from the analysis of generic 
supersymmetric $AdS_5$ solutions of type IIB supergravity 
given in~\cite{Gauntlett:2005ww}. The ten-dimensional metric, in
Einstein frame, is 
\bea\label{10dmetric}
g_{E}  =  \me^{2\Delta} \left(g_{{AdS}}  +  g_Y\right)~,
\eea
where $g_Y$ is a Riemannian  metric on the compact internal five-manifold $Y$, 
and $\Delta$ is a function on $Y$. The $AdS_5$ metric is normalized to
have unit radius, $R^{AdS}_{\mu\nu}=-4g^{AdS}_{\mu\nu}$. 
The remaining IIB fields are
the dilaton $\phi$, the NS three-form field strength $H=\dd B$, 
and the RR potentials $C=C_0+C_2+C_4$, with field strengths
$F=F_1+F_3+F_5=(\diff-H\wedge)C$. 
It is useful to introduce 
the complex combinations
\bea
P &=& \tfrac{\mathrm{i}}{2} \me^{\phi} F_1 + \tfrac{1}{2} \dd \phi
   ~, \nonumber \\
G &=& -\me^{-\phi/2}H-{\mathrm i}\me^{\phi/2}F_3~, 
\eea
together with $Q=-\tfrac{1}{2}\me^{\phi}F_1$. 
All of these fields are taken to be pull-backs of forms 
on $Y$, so as to preserve the $SO(4,2)$ symmetry, with the exception of 
the self-dual five-form $F_5$, which necessarily takes the form
\bea\label{5flux}
F_5  =  f\left(\vol_{{AdS}}  +  \vol_Y\right)~,
\eea
where $f$ is a constant. The conventions for the volume forms $\vol_{AdS}$ and $\vol_Y$ are
given in~\cite{Gauntlett:2005ww}. To match standard contact structure
conventions, it will be useful here to take
$\widetilde{\vol}_Y=-\vol_Y$ as defining our orientation for
integrating five-forms on $Y$.  

As explained in detail in \cite{Gauntlett:2005ww}, 
a supersymmetric $AdS_5$ solution is specified by 
two $Spin(5)$ spinors $\xi_1$, $\xi_2$ on $Y$ satisfying the 
following system of equations 
\bea
 0&=&  D_m \xi_1
      + \tfrac{\mathrm{i}}{4} \left(\me^{-4\Delta}f-2\right)\gamma_m \xi_1
      + \tfrac{1}{8} \me^{-2\Delta} G_{mnp}\gamma^{np}\xi_2
    \nonumber\\
  0&=&  \bar{D}_m\xi_2
      -  \tfrac{\mathrm{i}}{4} \left(\me^{-4\Delta}f+2\right)\gamma_m \xi_2
      + \tfrac{1}{8} \me^{-2\Delta} 
{}G_{mnp}^{*}\gamma^{np}\xi_1
      \nonumber\\
 0&=&  (\gamma^m\de_m\Delta-\tfrac{{\mathrm i}}{4}{f}\me^{-4\Delta}+{\mathrm i})\xi_1
      - \tfrac{1}{48}\me^{-2\Delta}\gamma^{mnp} {G}_{mnp} \xi_2
      \nonumber\\
 0&=&  (\gamma^m\de_m\Delta +\tfrac{{\mathrm i}}{4}f\me^{-4\Delta}+{\mathrm i})\xi_2
      - \tfrac{1}{48}\me^{-2\Delta}\gamma^{mnp}G_{mnp}^*\xi_1
      \nonumber\\
 0&=&   \gamma^m P_m \xi_2
      + \tfrac{1}{24} \me^{-2\Delta} \gamma^{mnp} G_{mnp} \xi_1
       \nonumber\\
 0&=&  \gamma^m P_m^* \xi_1
      + \tfrac{1}{24} \me^{-2\Delta} \gamma^{mnp} G_{mnp}^* \xi_2~.
       \label{SUSYeqns}
\eea
Here $\gamma_m$ generate the Clifford algebra for 
$(Y,g_Y)$, with $\gamma_{12345}=+1$, and 
$D_m=(\nabla_m-\tfrac{\mathrm{i}}{2}Q_m)$. 
The equations (\ref{SUSYeqns}) may be 
rewritten in terms of differential 
constraints on scalars, one-forms and two-forms that are constructed as bilinears in the 
spinors \cite{Gauntlett:2005ww}. We shall introduce and study two particular such one-forms 
in the next section.

\section{Contact structure}

We begin with the one-form bilinear 
\bea
K_{5\, m} \equiv \tfrac{1}{2}\left(
      \bar{\xi}_1\gamma_m\xi_1 + \bar{\xi}_2\gamma_m\xi_2\right)~.
\eea
Defining $\xi^m\equiv g_Y^{mn}K_{5\, n}$, it was  shown in \cite{Gauntlett:2005ww} that $\xi$
is Killing, preserves all of the bosonic fields, and is therefore
the canonical vector field dual to the $R$-symmetry of the ${\cal
  N}=1$ SCFT. 
We next consider the spinor bilinears 
\bea
 K_{4\, m} &\equiv& \tfrac{1}{2}\left(
      \bar{\xi}_1\gamma_m\xi_1 - \bar{\xi}_2\gamma_m\xi_2\right)~, \nonumber\\
 {\mathrm i} V_{mn} &\equiv& \tfrac{1}{2}\left(
      \bar{\xi}_1\gamma_{mn}\xi_1 - \bar{\xi}_2\gamma_{mn}\xi_2
      \right)~.    
\eea
Supersymmetry implies~\cite{Gauntlett:2005ww} that 
\be
 \me^{-4\Delta} \diff(\me^{4\Delta} K_4) = -2V~.
\ee
We claim that 
\be
\sigma \equiv \tfrac{4}{f} \me^{4\Delta} K_4
\ee
is a contact one-form on $Y$. 
In particular, one can readily
show that 
\be\label{volform}
\sigma \wedge \dd\sigma \wedge \dd\sigma 
   = \tfrac{128}{f^2} \me^{8\Delta} \widetilde\vol_Y~. 
\ee
Notice this calculation 
makes sense only if $f\neq0$, or equivalently $F_5\neq0$. (We shall
determine the quantization condition on $f$, which is related to the
D3-brane charge, in the next section.) With this assumption,
$\sigma\wedge\diff\sigma\wedge \diff\sigma$ is nowhere-zero, and so,
by definition, $\sigma$ is a contact form. 

Furthermore, again using the results of~\cite{Gauntlett:2005ww}, we
have 
\bea\label{Reeb}
1=\xi\lrcorner \sigma~,\qquad
0=\xi\lrcorner\diff\sigma~,
\eea
which shows that $\xi$ is also the unique ``Reeb vector field''
associated with the contact structure. 

If $(Y,\sigma)$ is a contact manifold, the product 
$X={\mathbb R}_{>0}\times Y$ has a natural symplectic structure with 
symplectic two-form, by definition closed and non-degenerate, 
\bea
\omega = \tfrac{1}{2}\dd ( r^2 \sigma)~,
\eea
where $r>0$ is a coordinate on ${\mathbb R}_{>0}$. 
Note that writing the unit $AdS_5$ metric in Poincar\'e coordinates,
the metric~(\ref{10dmetric}) can also be rewritten in the form of a
warped supersymmetric ${\mathbb R}^{3,1}\times X$ solution
\bea
g_E  =  \me^{2\Delta}r^2 g_{{\mathbb R}^{3,1}}  +  \me^{2\Delta}r^{-2}g_X~,
\eea
where $g_{\mathbb{R}_{3,1}}$ is the flat metric on
$\mathbb{R}^{3,1}$ and the six-dimensional metric on $X$ is given by
the cone metric $g_X =  \diff r^2  +  r^2g_Y$. In the $SE_5$ case,
$g_X$ is Calabi-Yau and $\omega$ is the 
corresponding K\"ahler form. 
In contrast, in the generic case, there is no longer such a simple
relation between $g_X$ and $\omega$:
in fact $X$ is a special kind of generalized complex geometry which will
be discussed in~\cite{GCGpaper}. 

\section{Central charge}

The central charge of the dual SCFT is related to the five-dimensional 
Newton constant $G_5$ \cite{kostas}. The latter, in turn, was computed in 
appendix E of \cite{Gauntlett:2005ww}, and is given by the supergravity formula
\bea\label{5dNewton}
G_5 = \frac{\kappa^2_{10}}{8\pi V_5}~, \quad \mbox{where}\quad V_5 \equiv \int_Y \me^{8\Delta}\widetilde\vol_Y~.
\eea
In string theory the five-form flux $F_5+H\wedge C_2=\dd C_4$
is quantized. Specifically, we have
\bea
N = 
\frac{1}{(2\pi l_s)^4g_s}\int_Y \left(F_5 + H \wedge C_2\right)~,
\eea
and after a calculation we find
\bea
N= -\frac{f}{(2\pi l_s)^4g_s}\int_Y \frac{1}{\sin^2 \zeta} \,
\widetilde\vol_Y \ , 
\label{fquant}
\eea
where \footnote{The $SE_5$ case is $\sin\zeta=1$, for which (\ref{fquant}) gives $f=4e^{4\Delta}=-(2\pi l_s)^4g_s N/\vol(Y)$, 
where $g_Y$ is normalized via 
$R^Y_{\mu\nu}=4g^Y_{\mu\nu}$.} $\sin\zeta \equiv \frac{1}{2}(\bar{\xi}_1\xi_1-\bar{\xi}_2\xi_2) = \frac{1}{4}f\me^{-4\Delta}$, 
with the last equality taken from \cite{Gauntlett:2005ww}. Here the
spinors are normalized so that
$\frac{1}{2}(\bar{\xi}_1\xi_1+\bar{\xi}_2\xi_2)=1$, as
in~\cite{Gauntlett:2005ww}.  
Combining (\ref{5dNewton}) and (\ref{fquant}) and using
$2\kappa_{10}^2=(2\pi)^7l_s^8g_s^2$ leads to
\bea
G_5 =\frac{8V_5}{\pi^2f^2N^2}~.
\eea
Finally, using equation (\ref{volform}) together with the formula for the central charge $a=\pi/8G_5$ and the fact that $a_{{\mathcal N}=4}=N^2/4$,
we arrive at the  remarkably simple formula (\ref{centralchargesigma}). 

\section{Some applications}

The formula (\ref{centralchargesigma}) has some immediate applications.

Suppose that the Reeb vector field $\xi$ is quasi-regular, meaning 
that its orbits all close. Its flow then defines a $U(1)$ action on $Y$, which is dual to a 
$U(1)$ $R$-symmetry (more generally the orbits of 
$\xi$ need not all close, as happens, for example, for infinitely many 
of the $Y^{pq}$ Sasaki-Einstein five-manifolds \cite{paper2}).  
Then, using formula (\ref{centralchargesigma}), 
we may prove that the $a$ central charge is always 
a rational number. Notice that this is predicted from the dual 
${\cal N}=1$ SCFT, using \reef{aandc} and the fact that the
$R$-charges are all rational numbers if the $R$-symmetry is $U(1)$.

To proceed, let $\chi$ denote the canonically normalized generator of the $U(1)$ action, 
so that $\xi=k\chi$ for some constant $k\in{\mathbb R}_{>0}$. 
Since  $\xi$ is nowhere-zero \cite{GCGpaper}, the $U(1)$ action
 generated by $\chi$ is locally free (all isotropy groups are finite), and the orbit space will in 
general be an orbifold $M$. Thus $Y$ is the total space of a $U(1)$ principal
orbibundle $\mathcal{L}$ over $M$. 
Then the relations (\ref{Reeb}) imply 
that $k\sigma$ is a connection one-form on this 
bundle. We thus compute
\bea
\frac{k^3}{(2\pi)^3}\int_Y \sigma\wedge \dd\sigma\wedge\dd\sigma= \int_M c_1^2({\cal L}) \in {\mathbb Q}~.
\eea
Here we have used the fact that the Chern numbers of an orbibundle 
are rational. Finally, the constant $k$ is also necessarily rational. 
This follows, for example, since the scalar $S\equiv \bar{\xi}_2^c\xi_1$ 
has charge $-3$ under $\xi$ \cite{Gauntlett:2005ww}, whereas 
it must have an integer charge $-n\in {\mathbb Z}$ under $\chi$; thus 
$k=3/n\in{\mathbb Q}$, and it follows from (\ref{centralchargesigma}) 
that $a\in{\mathbb Q}$.

In general, we may also rewrite (\ref{centralchargesigma}) in terms of the symplectic structure on $X$ as
\be\label{centralchargeomega}
\frac{a_{{\cal N}=4}}{a}  = \frac{1}{(2\pi)^3}\int_X \me^{-r^2/2}\, \frac{\omega^3}{3!}~.
\ee
This is a Duistermaat-Heckman integral, with Hamiltonian function
$\mathcal{H}=r^2/2$ for the Reeb vector field $\xi$, so $\dd
\mathcal{H}=-\xi\lrcorner\omega$. The power of the formula 
(\ref{centralchargeomega}) is that it may be evaluated 
using localization \cite{MSY2}. This is easiest to explain when the
solution is toric, 
meaning that there is a $U(1)^3$ action on $Y$ 
under which $\sigma$ (and hence $\omega$ under the lift to $X$) is
invariant. 
In this case there is a  
moment map $\mu:X\rightarrow {\mathbb R}^3$, whose image is a 
strictly convex rational polyhedral cone.
As explained in 
\cite{MSY2}, the integral (\ref{centralchargeomega}) may be evaluated
by taking any simplicial resolution $\mathcal{P}$ of this cone, and
evaluating 
\bea
\frac{1}{(2\pi)^3}\int_X \me^{-r^2/2}\, \frac{\omega^3}{3!} = 
\sum_{\mathrm{vertices\ }\mathrm{p}\in{\cal P}}  \prod_{i=1}^3 \frac{1}{\langle
\xi, u^{\mathrm p}_i\rangle}~,
\eea
where $u_i^{\mathrm p}$, $i=1,2,3$, are the three edge vectors of the
moment polytope $\mathcal{P}$ at the vertex point $\mathrm{p}$. The
vertices of $\mathcal{P}$  
correspond to the $U(1)^3$ fixed points of a symplectic toric resolution ${X}_{{\mathcal P}}$ of $X$.
Thus, remarkably, these results of \cite{MSY2} hold in general,
even when there are non-trivial fluxes turned on and $X$ is not
Calabi-Yau.

\section{Conformal dimension of BPS branes}

Static probe D3-branes that wrap supersymmetric three-submanifolds
$\Sigma_3\subset Y$ are dual to chiral primary operators in the dual SCFT. 
We claim that the conformal dimension $\Delta({\mathcal O}_{\Sigma_3})$ of 
the operator is given by (\ref{conformal}). To show this
we define $\Delta({\mathcal O}_{\Sigma_3})$ 
as the coefficient of the logarithmically divergent part of the action 
of a Euclidean D3-brane wrapped on $\Sigma_3$ and the $r$-direction; 
see, e.g., \cite{baryonpaper}. The wrapped brane is supersymmetric 
only if it satisfies the calibration condition that, restricting to the worldvolume ${\mathbb
  R}_{>0}\times\Sigma_3$, 
\bea\label{calibrate}
\frac{f}{8}\frac{\dd r}{r} \wedge\sigma\wedge \dd \sigma = \me^{-\phi}
\sqrt{\det (h-B)}\, \diff x_1\wedge\cdots\wedge\diff x_4
\eea
where $x_i$ are coordinates on the worldvolume and $h_{ij}$ is the
induced (string frame) metric. 
Up to a factor of the tension of the D3-brane, given by $1/(2\pi
l_s)^4g_s$, the term on the right is precisely the Dirac-Born-Infeld
Lagrangian, and gives 
the only contribution to the logarithmically
divergent part of the action (for further details see
\cite{GCGpaper}).   

Again, $\Delta({\mathcal O}_{\Sigma_3})$ is rational for solutions
with $U(1)_R$ symmetry, since $\Sigma_3$ fibres over a two-cycle
$\Sigma_2\subset M$, and the numerator of (\ref{conformal}) is then
$(2\pi)^3N/k^2$ times the Chern number
$\int_{\Sigma_2}c_1(\mathcal{L})$, which is  rational. This is as
expected, since in the dual SCFT $\Delta=\tfrac{3}{2}R$, where $R$ is
the $R$-charge. 

\section{Concluding remarks}

The fact that several key properties of $\mathcal{N}=1$ SCFTs are
determined by the $R$-symmetry of the theory should be reflected in
the geometry of the dual supergravity background. For the case of
generic $AdS_5\times Y$ solutions of type IIB supergravity, we have
shown that the relevant object is a canonical contact structure on
$Y$, determined by the Reeb vector field. This led to succinct
and universal 
formulae for calculating the central charge and certain conformal
dimensions, without knowing the full solution.  
Potentially, as in the $SE_5$ case, the contact structure is rich
enough to encode other properties of the field theory. For
instance, the principal of $a$-maximization~\cite{kenbrian} suggests
that the Reeb vector field itself might be determined from general
geometric data. For the class of type IIB solutions considered here,
it may be that knowing only the symplectic structure of the cone $X$,
together with some other (integral) data of the solution, analogous to
the results of \cite{MSY1,MSY2} for the special Sasaki-Einstein
case, is sufficient. 

The generic type IIB $AdS_5$ solutions considered here, like those
in~\cite{MSY1,MSY2}, have non-vanishing D3-brane charge. It is  
natural to wonder what happens for the special case where this is no
longer true (assuming that such solutions exist). The corresponding
formulae would have to be quite different since there is no longer a
contact structure. It is also known that there are rich classes of
supersymmetric $AdS_5\times M_6$ solutions of $D=11$ supergravity that
are dual to ${\cal N}=1$ SCFTs~\cite{Gauntlett:2004zh}. To obtain the
geometric formulae for this class will also require new ideas, since
now the cone geometry is seven-dimensional.


\medskip
\noindent
M.G. is supported by the Berrow Foundation, J.P.G. by an EPSRC Senior
Fellowship and a Royal Society Wolfson Award, E.P. by a STFC
Postdoctoral Fellowship and J.F.S. by a Royal Society University
Research Fellowship. 


\end{document}